\documentclass[aps,prb,twocolumn,groupedaddress,10pt]{revtex4-1}

\usepackage{amsmath,amssymb,amsthm,mathtools}
\usepackage{bm}

\newcommand{\ii}{\mathrm{i}}
\newcommand{\ee}{\mathrm{e}}
\newcommand{\vvv}[1]{\bm{#1}}
\newcommand{\curl}[1]{\nabla\times #1}
\newcommand{\pder}[2]{\frac{\partial #1}{\partial #2}}

\begin{document}

\title{Retrieval of all effective susceptibilities in nonlinear metamaterials}

\author{St\'ephane Larouche}
\email{stephane.larouche@northropgrummannext.net}
\affiliation{NG Next, Northrop Grumman Corporation, Redondo Beach, CA 90278, USA}
\author{Vesna Radisic}
\affiliation{NG Next, Northrop Grumman Corporation, Redondo Beach, CA 90278, USA}

\date{\today}

\begin{abstract}
Electromagnetic metamaterials offer a great avenue to engineer and amplify the nonlinear response of materials. Their electric, magnetic, and magneto-electric linear and nonlinear response are related to their structure, providing unprecedented liberty to control those properties. Both the linear and the nonlinear properties of metamaterials are typically anisotropic. While the methods to retrieve the effective linear properties are well established, existing nonlinear retrieval methods have serious limitations. In the present work, we generalize a nonlinear transfer matrix approach to account for all nonlinear susceptibility terms and show how to use this approach to retrieve all effective nonlinear susceptibilities of metamaterial elements. The approach is demonstrated using sum frequency generation, but can be applied to other second order or higher order processes.
\end{abstract}


\maketitle

\section{Introduction}

Nonlinear metamaterials offer unprecented control over the nonlinear properties of materials, as well as access to new exotic properties not available in standard materials~\cite{Shadrivov2015Nonlinear}. It has been proposed that they will play a significant role for new devices operating over a wide range of the electromagnetic spectrum~\cite{Zheludev2012From,Urbas2016Roadmap}.

The advantage of metamaterials over other structured materials is the possibility to assign them effective properties, i. e. the properties of an homogeneous effective material producing the same effect as the metamaterial. This is a powerful concept allowing one to separate the design of the metamaterial unit cell from that of the device where the metamaterial is used. The latter would be prohibitively difficult, if it were even possible, if one needed to consider all the details of the structure of every unit cell.

The retrieval of metamaterial effective linear properties is well established~\cite{Smith2002Determination, Chen2004Robust, Chen2005Retrieval}. Some approaches for the retrieval of effective nonlinear properties of metamaterials have also been proposed~\cite{Larouche2010Retrieval,Rose2010Nonlinear,Larouche2015Constitutive,Gorlach2016Nonlocal,Wang2017Retrieval}, but they are limited to metamaterials with enough symmetry (such as isotropic metamaterials) that it is possible to treat the elements of the nonlinear susceptibility tensors separately. Many nonlinear metamaterials do not respect this assumption and it is necessary to generalize the nonlinear retrieval approach to account for all nonlinear susceptibilities at once~\cite{Hooper2017Strong}.

In this paper, we present a general retrieval approach for nonlinear metamaterials. In order to perform the retrieval, we need to be able to calculate the fields in a slab of bi-anisotropic material. Therefore, we first review a transfer matrix approach for linear bi-anisotropic materials. Then we extend a nonlinear transfer matrix approach to include such materials, as well as magnetic and magneto-electric nonlinearities. Finally, we show how the nonlinear transfer matrix approach can be used in a retrieval approach to determine all the nonlinear susceptibility terms.

\section{Transfer matrix approach for bi-anisotropic materials}

Calculating the reflection and transmission of a stack of material slabs is a one-dimensional problem. For linear isotropic materials, the problem can be separated into transverse electric (TE) and transverse magnetic (TM) waves. Many approaches based on $2\times 2$ matrices exist for this problem, and are often presented in electromagnetics and optics textbooks, such as that of Born and Wolf~\cite{Born1999Principles}. In the case of anisotropic, magneto-electric, or bi-anisotropic materials, it is impossible to separate the problem into TE and TM waves. In this case, $4\times 4$ matrix approaches can be used~\cite{Berreman1972Optics, Yeh1979Electromagnetic}; in this paper, we use an approach similar to that of Berreman~\cite{Berreman1972Optics}. As this approach is not widely used, and it is necessary to understand it for the nonlinear calculations, we review it in this section. We first determine the propagation modes in a bi-anisotropic material. We then build transfer and scattering matrices to calculate reflection and transmission, as well as the fields at all positions in the stack.

\subsection{Propagation modes}

In the absence of free charges and currents, Maxwell's curl equations in the frequency domain ($\ee^{-\ii\omega t}$ time convention, where $\omega$ is the angular frequency and $t$ the time) are
\begin{subequations}
\begin{align}
	\curl{\vvv{E}}       &= \ii\omega\vvv{B},\text{ and}\\
	-\curl{\vvv{H}}      &= \ii\omega\vvv{D},
\end{align}
\end{subequations}
where $\vvv{E}$ and $\vvv{H}$ are the electric and magnetic fields, $\vvv{B}$ is the magnetic induction, and $\vvv{D}$ is the electric displacement. The material equations are 
\begin{subequations}
\begin{align}
	\vvv{D} &= \epsilon\vvv{E} + \xi\vvv{H},\text{ and}\\
	\vvv{B} &= \zeta\vvv{E}    + \mu\vvv{H},
\end{align}
\end{subequations}
where $\epsilon$ and $\mu$ are the permittivity and the permeability, while $\xi$ and $\zeta$ are the magneto-electric coupling coefficients. All the material properties are $3\times 3$ rank two tensors.

Maxwell's equations and the material equations can be combined in a single matrix equation,
\begin{equation}
	\begin{bmatrix}
		[0]       & -[\curl{}]\\
		[\curl{}] & [0]
	\end{bmatrix}
	\begin{bmatrix}
		\vvv{E}\\
		\vvv{H}
	\end{bmatrix}
	= \ii\omega\begin{bmatrix}
		[\epsilon] & [\xi]\\
		[\zeta]    & [\mu]
	\end{bmatrix}
	\begin{bmatrix}
	      \vvv{E}\\
	      \vvv{H}
        \end{bmatrix},
	\label{EqWaveEquation}
\end{equation}
where $[0]$ is a $3\times 3$ null matrix and
\begin{equation*}
	[\curl{}]
	= \begin{bmatrix}
		          0 & -\pder{}{z} &  \pder{}{y}\\
		 \pder{}{z} &           0 & -\pder{}{x}\\
		-\pder{}{y} &  \pder{}{x} &          0
	\end{bmatrix}.
\end{equation*}
Inside a uniform medium, this equation is a first-order wave equation with solutions of the form
\begin{equation}
	\begin{bmatrix}
	      \vvv{E}(x,y,z)\\
	      \vvv{H}(x,y,z)
        \end{bmatrix}
	= \exp(\ii(k_x x+k_y y+k_z z))
	\begin{bmatrix}
		\vvv{E}(0)\\
		\vvv{H}(0)
        \end{bmatrix},
\end{equation}
where $k_x$, $k_y$, and $k_z$ are the components of the propagation constant. Therefore the partial derivatives $\pder{}{x,y,z} $ can be replaced by $\ii k_{x,y,z}$ and the curl operator by
\begin{equation*}
	[\curl{}]
	= \begin{bmatrix}
		       0 & -\ii k_z &  \ii k_y \\
		 \ii k_z &        0 & -\ii k_x \\
		-\ii k_y &  \ii k_x &        0
	\end{bmatrix}.
\end{equation*}

\begin{figure}
	\includegraphics{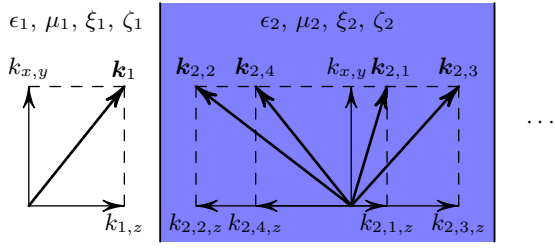}
	\caption{(Color online) The system considered consists of a series of slabs of potentially bi-anisotropic materials. The in-plane propagation constants, $k_x$ and $k_y$, are imposed by the applied waves (in medium 1 here), and invariant throughout the whole system. Inside every slab, there are four modes with potentially different propagation constants.}
	\label{FigMultislab}
\end{figure}

We are looking for a method to calculate the properties of a stack of material slabs. Without loss of generality, let us suppose that the material properties only depend on $z$, while they are uniform in the $xy$ plane as illustrated in Fig.~\ref{FigMultislab}. The system consists of a series of slabs between two semi-infinite media. Waves can be applied on either sides of the stack. Since all the elements of the system are uniform in the $xy$ plane, $k_x$ and $k_y$ remain invariant throughout all the slabs and are known from the applied wave. The propagation constant in the $z$ direction, $k_z$, varies with the material properties and needs to be determined, as well as the magnitude and orientation of the electric and magnetic fields.

The third and sixth rows of Eq.~\ref{EqWaveEquation} are
\begin{multline}
	\begin{bmatrix}
		       0 &        0 & 0 &  \ii k_y & -\ii k_x & 0 \\
		-\ii k_y &  \ii k_x & 0 &        0 &        0 & 0 \\
	\end{bmatrix}
	\begin{bmatrix}
	      \vvv{E}\\
	      \vvv{H}
        \end{bmatrix}\\
	= \ii\omega\begin{bmatrix}
		\epsilon_{zx} & \epsilon_{zy} & \epsilon_{zz} & \xi_{zx} & \xi_{zy} & \xi_{zz}\\
		   \zeta_{zx} &    \zeta_{zy} &    \zeta_{zz} & \mu_{zx} & \mu_{zy} & \mu_{zz}\\
	\end{bmatrix}
	\begin{bmatrix}
	      \vvv{E}\\
	      \vvv{H}
        \end{bmatrix}.
\end{multline}
These two equations relate the components of the fields when $k_x$, $k_y$, and the material properties are known. They can be used to express two field components as a function of the four others. Since our goal is to develop an approach for multilayer stacks, it is obviously advantageous to work with the components of the fields that are continuous at the interface between the layers, namely $E_x$, $E_y$, $H_x$, and $H_y$, while $E_z$ and $H_z$ are eliminated. The $z$ components of the fields are
\begin{multline}
	\begin{bmatrix}
		E_z\\
		H_z
	\end{bmatrix}
	= -\begin{bmatrix}
		\epsilon_{zz} & \xi_{zz}\\
		   \zeta_{zz} & \mu_{zz}\\
	\end{bmatrix}^{-1}\\
	\begin{bmatrix}
		\epsilon_{zx}                    & \epsilon_{zy}                    &  \xi_{zx}-\frac{k_y}{\omega} & \xi_{zy}+\frac{k_x}{\omega} \\
		   \zeta_{zx}+\frac{k_y}{\omega} &    \zeta_{zy}-\frac{k_x}{\omega} &  \mu_{zx}                    & \mu_{zy}                    \\
	\end{bmatrix}
	\begin{bmatrix}
		E_x\\
		E_y\\
		H_x\\
		H_y
	\end{bmatrix}.
	\label{EqZComponents}
\end{multline}
Separating the $z$ components in the four remaining equations yields
\begin{multline*}
	\begin{bmatrix}
		       0 &        0 &        0 &  \ii k_z \\
		       0 &        0 & -\ii k_z &        0 \\
		       0 & -\ii k_z &        0 &        0 \\
		 \ii k_z &        0 &        0 &        0 \\
	\end{bmatrix}
	\begin{bmatrix}
		E_x\\
		E_y\\
		H_x\\
		H_y
	\end{bmatrix}
	+	\begin{bmatrix}
		       0 & -\ii k_y \\
		       0 &  \ii k_x \\
		 \ii k_y &        0 \\
		-\ii k_x &        0 \\
	\end{bmatrix}
	\begin{bmatrix}
		E_z\\
		H_z
	\end{bmatrix}\\
	= \ii\omega\left(\begin{bmatrix}
		\epsilon_{xx} & \epsilon_{xy} & \xi_{xx} & \xi_{xy} \\
		\epsilon_{yx} & \epsilon_{yy} & \xi_{yx} & \xi_{yy} \\
		   \zeta_{xx} &    \zeta_{xy} & \mu_{xx} & \mu_{xy} \\
		   \zeta_{yx} &    \zeta_{yy} & \mu_{yx} & \mu_{yy} \\
	\end{bmatrix}
	\begin{bmatrix}
		E_x\\
		E_y\\
		H_x\\
		H_y
	\end{bmatrix}
	+ \begin{bmatrix}
		\epsilon_{xz} & \xi_{xz} \\
		\epsilon_{yz} & \xi_{yz} \\
		   \zeta_{xz} & \mu_{xz} \\
		   \zeta_{yz} & \mu_{yz} \\
	\end{bmatrix}
	\begin{bmatrix}
		E_z\\
		H_z
	\end{bmatrix}\right)
\end{multline*}
or, moving all $z$ components to the right side, inserting the solution from Eq.~\ref{EqZComponents}, eliminating the matrix on the left side, reordering the rows, and taking the inverse of $H_x$ (to make its value positive for forward propagation in a right-handed material),
\begin{equation}
	k_z\begin{bmatrix}
		 E_x\\
		 H_y\\
		 E_y\\
		-H_x
	\end{bmatrix}
	= \Delta
	\begin{bmatrix}
		 E_x\\
		 H_y\\
		 E_y\\
		-H_x
	\end{bmatrix},
	\label{EqEigenequation}
\end{equation}
where
\begin{widetext}
\begin{multline}
	\Delta
	= \omega\begin{bmatrix}
		 0 &  0 &  0 &  1 \\
		 1 &  0 &  0 &  0 \\
		 0 &  0 & -1 &  0 \\
		 0 &  1 &  0 &  0 \\
	\end{bmatrix}
	\left(\begin{bmatrix}
		\epsilon_{xx} & \epsilon_{xy} & \xi_{xx} & \xi_{xy} \\
		\epsilon_{yx} & \epsilon_{yy} & \xi_{yx} & \xi_{yy} \\
		   \zeta_{xx} &    \zeta_{xy} & \mu_{xx} & \mu_{xy} \\
		   \zeta_{yx} &    \zeta_{yy} & \mu_{yx} & \mu_{yy} \\
	\end{bmatrix}
	\right.\\
	\left.
	- \begin{bmatrix}
		\epsilon_{xz}                    & \xi_{xz}+\frac{k_y}{\omega} \\
		\epsilon_{yz}                    & \xi_{yz}-\frac{k_x}{\omega} \\
		   \zeta_{xz}-\frac{k_y}{\omega} & \mu_{xz}                    \\
		   \zeta_{yz}+\frac{k_x}{\omega} & \mu_{yz}                    \\
	\end{bmatrix}
	\begin{bmatrix}
		\epsilon_{zz} & \xi_{zz}\\
		   \zeta_{zz} & \mu_{zz}\\
	\end{bmatrix}^{-1}
	\begin{bmatrix}
		\epsilon_{zx}                    & \epsilon_{zy}                    &  \xi_{zx}-\frac{k_y}{\omega} & \xi_{zy}+\frac{k_x}{\omega} \\
		   \zeta_{zx}+\frac{k_y}{\omega} &    \zeta_{zy}-\frac{k_x}{\omega} &  \mu_{zx}                    & \mu_{zy} \\
	\end{bmatrix}\right)
	\begin{bmatrix}
		 1 &  0 &  0 &  0 \\
		 0 &  0 &  1 &  0 \\
		 0 &  0 &  0 & -1 \\
		 0 &  1 &  0 &  0 \\
	\end{bmatrix}.
\end{multline}
\end{widetext}
Equation~\ref{EqEigenequation} is an eigenvalue equation. The propagation constants in the $z$ direction are the eigenvalues of $\Delta$ while the propagation modes are its eigenvectors. There are four propagation modes, each with its associated propagation constant.

It is worth noting that in anisotropic materials the direction of $k_z$ is not always the propagation direction. The latter must be determined using the Poynting vector, which can be calculated from the modes. There are always two forward and two backward propagating modes. The modes can also be grouped by polarization as there are always pairs of forward and backward propagating modes with the same polarization, such as right and left elliptically polarized.

In materials with high symmetry, the propagation constants are often degenerate. For example, in an isotropic material, all four propagation constants have the same absolute value, two of them positive and two of them negative. Furthermore, if the plane of incidence is either the $xz$ or $yz$ planes, $\Delta$ is block diagonal and can be separated into a pair of $2\times 2$ matrices for TE and TM waves.

\subsection{Transfer matrices}

In the previous section, we determined the propagation modes and constants in a uniform medium. In this section, we show how to calculate the properties of a stack of material slabs.

As mentioned in the previous section, there are four propagation modes in each slab of material. We can express the fields in a slab as a vector of the amplitudes of those four modes,
\begin{equation*}
	\vvv{A}
	= \begin{bmatrix}
		A_1\\
		A_2\\
		A_3\\
		A_4\\
	\end{bmatrix}.
\end{equation*}
The modes can be arranged in any order. For simplicity, let us suppose that modes 1 and 3 are forward propagating, while modes 2 and 4 are backward propagating. Let us also suppose that modes 1 and 2, as well as modes 3 and 4, are of the same polarization pairwise.

At the interface between two materials, in the absence of surface currents and charges, the tangential components of the electric and magnetic fields are continuous. To determine the tangential components at the interface, we need to sum the $x$ and $y$ components from the four modes using
\begin{equation*}
	\begin{bmatrix}
		 E_x\\
		 H_y\\
		 E_y\\
		-H_x
	\end{bmatrix}
	= \Pi \vvv{A},
\end{equation*}
where $\Pi$ is a matrix whose rows are the propagation modes inside of the slab (in the same order selected for $\vvv{A}$). At the interface between slabs $i-1$ and $i$,
\begin{equation*}
	\Pi_{i-1}\vvv{A}_{i-1}(z_i) = \Pi_i\vvv{A}_i(z_i),
\end{equation*}
where $z_i$ is the position of the interface. Therefore, the transfer matrix of the interface is
\begin{equation}
	M_{i-1,i} = \Pi_{i}^{-1}\Pi_{i-1}.
\end{equation}

Inside a uniform slab, the modes propagate without interacting and simply accumulate phase. The transfer matrix related to the propagation inside slab $i$ is
\begin{equation}
	\Phi_i
	= \begin{bmatrix}
		\ee^{\ii k_{z,i,1}d_i} &                      0 &                      0 &                      0 \\
		                     0 & \ee^{\ii k_{z,i,2}d_i} &                      0 &                      0 \\
		                     0 &                      0 & \ee^{\ii k_{z,i,3}d_i} &                      0 \\
		                     0 &                      0 &                      0 & \ee^{\ii k_{z,i,4}d_i} \\
	\end{bmatrix},
\end{equation}
where the $k_{z,i}$s are the eigenvalues of $\Delta_i$ and $d_i = z_{i+1}-z_i$ is the thickness of the slab.

The transfer matrix of a whole stack of slabs is obtained by multiplying the individual transfer matrices. For example, the transfer matrix of a slab of material $2$ sandwiched between two semi-infinite media $1$ and $3$ is
\begin{equation}
	M = M_{2,3}\Phi_2M_{1,2}.
\end{equation}

\subsection{Scattering matrices (reflection and transmission)}

The mode amplitudes on both sides of a stack of $n-2$ slabs are related by the transfer matrix of the system such that
\begin{equation}
	\vvv{A}_n = M\vvv{A}_1.
\end{equation}
However, in practice, it is rare that all the amplitudes on either side are known. More often, the incoming waves are known, while the outgoing waves need to be determined. The previous equation can be expressed as
\begin{equation*}
	\begin{bmatrix}
		A_{n,1}\\
		A_{n,2}\\
		A_{n,3}\\
		A_{n,4}\\
	\end{bmatrix}
	= M\begin{bmatrix}
		A_{1,1}\\
		A_{1,2}\\
		A_{1,3}\\
		A_{1,4}\\
	\end{bmatrix}.
\end{equation*}
According to the convention established in the previous section, the applied waves $A_{1,1}$, $A_{1,3}$, $A_{n,2}$, and $A_{n,4}$ are known while the outgoing waves $A_{1,2}$, $A_{1,4}$, $A_{n,1}$, and $A_{n,3}$ are not. By reorganizing the system of equations, it is straightforward to demonstrate that
\begin{equation}
	\begin{bmatrix}
		A_{1,2}\\
		A_{1,4}\\
		A_{n,1}\\
		A_{n,3}\\
	\end{bmatrix}
	= S\begin{bmatrix}
		A_{1,1}\\
		A_{1,3}\\
		A_{n,2}\\
		A_{n,4}\\
	\end{bmatrix},
\end{equation}
where
\begin{equation}
	S
	= -\begin{bmatrix}
		M_{12} & & M_{14} & -1 &  0 \\
		M_{22} & & M_{24} &  0 &  0 \\
		M_{32} & & M_{34} &  0 & -1 \\
		M_{42} & & M_{44} &  0 &  0 \\
	\end{bmatrix}^{-1}
	\begin{bmatrix}
		M_{11} & & M_{13} &  0 &  0 \\
		M_{21} & & M_{23} & -1 &  0 \\
		M_{31} & & M_{33} &  0 &  0 \\
		M_{41} & & M_{43} &  0 & -1 \\
	\end{bmatrix}
	\label{EqScatteringMatrix}
\end{equation}
is the scattering matrix.

The elements of the scattering matrix can be seen as reflection and transmission coefficients. For general bi-anisotropic material, it is impossible to get a single value for the transmission and the reflection since both of them can occur in two modes.

Knowing the amplitude of all the modes on either sides of the system, interface and propagation transfer matrices can be used to calculate the amplitudes at any position inside the system. We use these in the next section to calculate the nonlinear effects.

Note that the scattering matrix Eq.~\ref{EqScatteringMatrix} relates the amplitudes of modes defined by their tangential components. This is different from the amplitudes determined by many simulation software, or experimentally, which usually include out-of-plane components as well. The full modes can be calculated using Eq.~\ref{EqZComponents} and the scattering matrix renormalized accordingly.

\section{Nonlinear transfer matrix approach}

In the previous section, we established how to calculate the fields inside a stack of bi-anisotropic linear materials. In this section, we extend the analysis to nonlinear effects. Our approach generalizes that of Bethune\cite{Bethune1989Optical,Bethune1991Optical}, who only considered the case of a simply anisotropic material with electric nonlinearity.

As is customary, we assume that the material nonlinearity can be described using a power series expansion. For simplicity, let us consider a second order process, sum frequency generation (SFG). In SFG, the interaction of two applied waves at $\omega_1$ and $\omega_2$ produces a wave at $\omega_3 = \omega_1+\omega_2$. This case can easily be generalized to difference frequency generation by changing the signs of the frequency and the propagation vectors at either $\omega_1$ or $\omega_2$, and to second harmonic generation (SHG) by considering $\omega_1 = \omega_2$.

In homogeneous materials, only electric nonlinearities are typically present. However, in metamaterials, magnetic and magneto-electric nonlinearities are often present, and dominate in many geometries. It is even quite frequent to have multiple nonlinearities present in the same metamaterial and they must therefore all be considered at once.

To represent all possible quadratic nonlinear effects, we use the notation $\chi_{abc,\alpha\beta\gamma}^{(2)}$ where $a$, $b$, and $c$ can each take the values e or m while $\alpha$, $\beta$, and $\gamma$ can each take the values $x$, $y$, or $z$. The first pair of indices, $a$ and $\alpha$, represent the nature (electric or magnetic) and orientation of the field at the nonlinear frequency, while the two other pairs indicate the same properties for the two applied waves. For example, $\chi_{\text{emm},xyy}^{(2)}$ indicates how the $y$ component of the magnetic fields at $\omega_1$ and $\omega_2$ interact to create a nonlinear polarization in the $x$ axis. Considering the electric or magnetic nature of the phenomenon at all three frequencies involved, and the three space dimensions, there is a total of $2^33^3 = 216$ terms in the second order susceptibility tensors.

We assume that the nonlinear process is weak enough that it does not significantly affect the amplitude of the applied waves. This is known as the non-depleted pump approximation, and is valid in many practical applications of nonlinear materials. In the particular case of retrieval, which is our main interest here, it is always possible to control the amplitude of the applied waves such that the approximation is respected. With this approximation, the nonlinear wave can be calculated by following a series of simple steps: (1) determine the amplitude of the electric and magnetic fields at $\omega_1$ and $\omega_2$ at all positions inside the stack using the linear transfer matrix approach described in previous sections; (2) calculate the nonlinear polarization and magnetization, and bound waves at $\omega_3$; and (3) couple bound waves to propagating waves at $\omega_3$. We now show how to perform steps 2 and 3.

\subsection{Nonlinear polarization and magnetization, and bound waves}

The interaction between waves at $\omega_1$ and $\omega_2$ propagating with $\vvv{k}(\omega_1)$ and $\vvv{k}(\omega_2)$ creates nonlinear polarization and magnetization at $\vvv{k}^{\text{NL}} = \vvv{k}(\omega_1)+\vvv{k}(\omega_2)$. This must be distinguished from normal waves propagating at $\omega_3$ which preserve the in-plane components of the propagation vector according to $k_{x,y}(\omega_3) = k_{x,y}(\omega_1)+k_{x,y}(\omega_2)$, but where the $z$ component must be determined according to the properties of the material with Eq.~\ref{EqEigenequation}.

The propagation constants $\vvv{k}(\omega_1)$ and $\vvv{k}(\omega_2)$ can both take four different values, so $\vvv{k}^{\text{NL}}$ can take 16 (possibly degenerate) values. In the non-depleted pump approximation, the different components do not interact. Therefore, they can be calculated separately and then simply added. At any position in the system, the nonlinear polarization and magnetization for one of those components are
\begin{widetext}
\begin{subequations}
\label{EqNonlinearPolarizationAndMagnetization}
\begin{align}
	P_{\vvv{k}_{p,q}^{\text{NL}},\alpha}^{\text{NL}}
	& = \begin{multlined}[t]
	    \epsilon_0\sum_{jk} \chi_{\text{eee},\alpha\beta\gamma}^{(2)}(\omega_3;\omega_1,\omega_2)E_{\vvv{k}_p(\omega_1),\beta}(\omega_1)E_{\vvv{k}_q(\omega_2),\gamma}(\omega_2)
	                       +\chi_{\text{eem},\alpha\beta\gamma}^{(2)}(\omega_3;\omega_1,\omega_2)E_{\vvv{k}_p(\omega_1),\beta}(\omega_1)H_{\vvv{k}_q(\omega_2),\gamma}(\omega_2)\\
	                       +\chi_{\text{eme},\alpha\beta\gamma}^{(2)}(\omega_3;\omega_1,\omega_2)H_{\vvv{k}_p(\omega_1),\beta}(\omega_1)E_{\vvv{k}_q(\omega_2),\gamma}(\omega_2)
	                       +\chi_{\text{emm},\alpha\beta\gamma}^{(2)}(\omega_3;\omega_1,\omega_2)H_{\vvv{k}_p(\omega_1),\beta}(\omega_1)H_{\vvv{k}_q(\omega_2),\gamma}(\omega_2),
			\end{multlined}\\
	M_{\vvv{k}_{p,q}^{\text{NL}},\alpha}^{\text{NL}}
	& = \begin{multlined}[t]
	    \mu_0\sum_{jk} \chi_{\text{mee},\alpha\beta\gamma}^{(2)}(\omega_3;\omega_1,\omega_2)E_{\vvv{k}_p(\omega_1),\beta}(\omega_1)E_{\vvv{k}_q(\omega_2),\gamma}(\omega_2)
	  	              +\chi_{\text{mem},\alpha\beta\gamma}^{(2)}(\omega_3;\omega_1,\omega_2)E_{\vvv{k}_p(\omega_1),\beta}(\omega_1)H_{\vvv{k}_q(\omega_2),\gamma}(\omega_2)\\
		                +\chi_{\text{mme},\alpha\beta\gamma}^{(2)}(\omega_3;\omega_1,\omega_2)H_{\vvv{k}_p(\omega_1),\beta}(\omega_1)E_{\vvv{k}_q(\omega_2),\gamma}(\omega_2)
		                +\chi_{\text{mmm},\alpha\beta\gamma}^{(2)}(\omega_3;\omega_1,\omega_2)H_{\vvv{k}_p(\omega_1),\beta}(\omega_1)H_{\vvv{k}_q(\omega_2),\gamma}(\omega_2),
			\end{multlined}
\end{align}
\end{subequations}
\end{widetext}
where $\alpha$, $\beta$, and $\gamma$ take values $x$, $y$, and $z$, while $p$ and $q$ take values 1--4.

The nonlinear polarization and magnetization drive bound waves. Their amplitude can be determined by solving the wave equation in the presence of a driving term,
\begin{multline}
	\begin{bmatrix}
		      [0] & -[\curl{}]\\
		[\curl{}] &        [0]
	\end{bmatrix}_{\vvv{k}_{p,q}^{\text{NL}}}
	\begin{bmatrix}
		\vvv{E}_s\\
		\vvv{H}_s
	\end{bmatrix}_{\vvv{k}_{p,q}^{\text{NL}}}\\
	= \ii\omega_3\left(\begin{bmatrix}
		[\epsilon(\omega_3)] & [\xi(\omega_3)]\\
		[\zeta(\omega_3)]    & [\mu(\omega_3)]
	\end{bmatrix}
	\begin{bmatrix}
	      \vvv{E_s}\\
	      \vvv{H_s}
   \end{bmatrix}_{\vvv{k}_{p,q}^{\text{NL}}}
	+\begin{bmatrix}
		\vvv{P}^{\text{NL}}\\
		\vvv{M}^{\text{NL}}
   \end{bmatrix}_{\vvv{k}_{p,q}^{\text{NL}}}
	\right),
	\label{EqWaveEquationS}
\end{multline}
where the curl matrix includes the propagation constants used when calculating the nonlinear polarization and magnetization, while the material matrix is calculated at $\omega_3$. Therefore, the bound waves are
\begin{multline}
	\begin{bmatrix}
		\vvv{E}_s\\
		\vvv{H}_s
	\end{bmatrix}_{\vvv{k}_{p,q}^{\text{NL}}}
	= \left(\frac{-\ii}{\omega_3}\begin{bmatrix}
		      [0] & -[\curl{}]\\
		[\curl{}] &        [0]
	\end{bmatrix}_{\vvv{k}_{p,q}^{\text{NL}}}
	\right.\\
	\left.
	- \begin{bmatrix}
		[\epsilon(\omega_3)] & [\xi(\omega_3)]\\
		[\zeta(\omega_3)]    & [\mu(\omega_3)]
	\end{bmatrix}
	\right)^{-1}
	\begin{bmatrix}
		\vvv{P}^{\text{NL}}\\
		\vvv{D}^{\text{NL}}
  \end{bmatrix}_{\vvv{k}_{p,q}^{\text{NL}}}.
\end{multline}
If the material properties are the same at all frequencies involved (or more precisely if the modes are the same), the above system of equations is singular. This corresponds to perfect phase matching. This singularity can be removed when the bound waves included in Eq.~\ref{EqS} below. In a numerical implementation, one can simply impose a small difference.

\subsection{Coupling to propagating waves}

Now that we have determined the bound waves created by the nonlinear process, we must determine how they couple to propagating waves. As in the linear case, the in-plane components of the waves must be continuous at the interface between layers. In the non-depleted pump approximation, the nonlinearity of each layer can be treated separately and simply added. If layer $i$ is nonlinear, at its two interfaces,
\begin{align*}
	\Pi_{i-1}A_{i-1}(z_i)
	= \Pi_iA_i(z_i)
	+ \Pi_s\sum_{p,q}\begin{bmatrix}
	                   \vvv{E}_s(z_{i})\\
		                 \vvv{H}_s(z_{i})
	                 \end{bmatrix}_{\vvv{k}_{p,q}^{\text{NL}}}\text{ and}\\
	\Pi_iA_i(z_{i+1})
	+ \Pi_s\sum_{p,q}\begin{bmatrix}
		                 \vvv{E}_s(z_{i+1})\\
		                 \vvv{H}_s(z_{i+1})
	                 \end{bmatrix}_{\vvv{k}_{p,q}^{\text{NL}}}
	= \Pi_{i+1}A_{i+1}(z_{i+1})
\end{align*}
where all the matrices are implicitly calculated at $\omega_3$ and
\begin{equation}
	\Pi_s
	= \begin{bmatrix}
		   1 &  0 &  0 &  0 &  0 &  0 \\
		   0 &  0 &  0 &  0 &  1 &  0 \\
		   0 &  1 &  0 &  0 &  0 &  0 \\
		   0 &  0 &  0 & -1 &  0 &  0 \\
	  \end{bmatrix}
\end{equation}
is a matrix that selects the in-plane components of the bound waves.

Using $A_i(z_{i+1}) = \Phi_iA_i(z_i)$ and
\begin{equation*}
	\begin{bmatrix}
		\vvv{E}_s(z_{i+1})\\
		\vvv{H}_s(z_{i+1})
	\end{bmatrix}_{\vvv{k}_{p,q}^{\text{NL}}}
	= \exp(\ii{}k_{p,q,z}^{\text{NL}}d_i)
	  \begin{bmatrix}
		  \vvv{E}_s(z_{i})\\
		  \vvv{H}_s(z_{i})
	  \end{bmatrix}_{\vvv{k}_{p,q}^{\text{NL}}},
\end{equation*}
we can show that
\begin{equation}
	\vvv{A}_{i+1} = M_{i,i+1}\Phi_i(M_{i-1,i}\Phi_{i-1}\vvv{A}_{i-1}+\vvv{S}_i),
\end{equation}
where all the amplitudes vectors are evaluated at the bottom of their respective layers,
\begin{multline}
	\vvv{S}_i
	= \Phi_i^{-1}M_{s,i}
	  \sum_{p,q}\exp(\ii{}k_{p,q,z}^{\text{NL}}d)\begin{bmatrix}
		            \vvv{E}_s\\
		            \vvv{H}_s
	            \end{bmatrix}_{\vvv{k}_{p,q}^{\text{NL}}}\\
	  - M_{s,i}\sum_{p,q}\begin{bmatrix}
		           \vvv{E}_s\\
		           \vvv{H}_s
	           \end{bmatrix}_{\vvv{k}_{p,q}^{\text{NL}}},
	\label{EqS}
\end{multline}
and $M_{s,i} = \Pi_i^{-1}\Pi_s$. Isolating $\vvv{S}_i$ yields
\begin{equation}
	\vvv{S}_i = \Phi_i^{-1}M_{i,i+1}^{-1}\vvv{A}_{i+1} - M_{i-1,i}\Phi_{i-1}\vvv{A}_{i-1}.
\end{equation}
The amplitude vectors $\vvv{A}_{i+1}$ and $\vvv{A}_{i-1}$ can be related to the amplitude vectors in the semi-infinite media on either sides of the stack such that
\begin{equation}
	R^{-1}\vvv{A}_n - L\vvv{A}_1 = \vvv{S}_i,
	\label{EqSAndA}
\end{equation}
where
\begin{align}
	L &= M_{i-1,i}\cdots \Phi_2M_{1,2}\text{ and}\\
	R &= M_{n-1,n}\Phi_n\cdots M_{i,i+1}\Phi_i
\end{align}
are the transfer matrices on the left and right of layer $i$. In the particular case of a single layer, which is the one appropriate for retrieval, $L = M_{1,2}$ and $R = M_{2,3}\Phi_2$.

Finally, since there is no applied wave at $\omega_3$, only the outgoing waves need to be determined and Eq.~\ref{EqSAndA} simplifies to
\begin{equation}
	\begin{bmatrix}
		A_{1,2}\\
		A_{1,4}\\
		A_{n,1}\\
		A_{n,3}\\
	\end{bmatrix}
	= \left(
	   -L \begin{bmatrix}
	       0 & 0\\
	       1 & 0\\
	       0 & 0\\
	       0 & 1\\
			 \end{bmatrix}
		-R^{-1}\begin{bmatrix}
	           1 & 0\\
	           0 & 0\\
	           0 & 1\\
	           0 & 0\\
			     \end{bmatrix}
		\right)^{-1}\vvv{S}_i.
\end{equation}

\section{Retrieval approach}

In the previous section, we showed how to calculate the fields generated by a stack of nonlinear materials. In this section, we present an approach to perform the inverse operation: determine the unknown nonlinear susceptibilities of a slab of material. This approach applies to any nonlinear material, but it is of particular interest for metamaterials, which often are bi-anisotropic and have more than one nonlinear susceptibility. Our approach generalizes that of Larouche, Rose, and Smith~\cite{Larouche2015Constitutive}, who treated the simpler case of materials with separable nonlinear properties for each axis and polarization.

Looking at Eq.~\ref{EqNonlinearPolarizationAndMagnetization}, it is obvious that, while the nonlinear polarization and magnetization depend nonlinearly on the fields, they depend linearly on the nonlinear susceptibilities. This suggests that it is possible to build a linear system of equations to determine the unknown nonlinear susceptibilities. In the case of SFG, there is a total of 216 complex nonlinear susceptibilities. To determine all of them, it is necessary to build a system of 216 complex equations.

This system of equations must be built using non-redundant illumination conditions. Such conditions are obtained by changing the angle of incidence of the applied waves, their polarization, and whether they come from medium $1$, medium $n$, or both. To probe all nonlinearities, it is essential to use applied waves at $\omega_1$ and $\omega_2$ with all possible combinations of polarizations as well as oblique incidence. 

Each illumination condition generates four measurements: the amplitude of the two outgoing wave modes in media $1$ and $n$. Therefore, it is necessary to find 54 independent illumination conditions. With these 54 illumination conditions, we build a linear system of equations
\begin{multline}
	\begin{bmatrix}
		A_{1,2}\rvert_{\chi_{\text{eee},xxx}^{(2)}=1}^{\text{ill. cond. 1}} & A_{1,2}\rvert_{\chi_{\text{eee},xxy}^{(2)}=1}^{\text{ill. cond. 1}} & \cdots \\
		A_{1,4}\rvert_{\chi_{\text{eee},xxx}^{(2)}=1}^{\text{ill. cond. 1}} & A_{1,2}\rvert_{\chi_{\text{eee},xxy}^{(2)}=1}^{\text{ill. cond. 1}} & \cdots \\
		A_{n,1}\rvert_{\chi_{\text{eee},xxx}^{(2)}=1}^{\text{ill. cond. 1}} & A_{1,2}\rvert_{\chi_{\text{eee},xxy}^{(2)}=1}^{\text{ill. cond. 1}} & \cdots \\
		A_{n,3}\rvert_{\chi_{\text{eee},xxx}^{(2)}=1}^{\text{ill. cond. 1}} & A_{1,2}\rvert_{\chi_{\text{eee},xxy}^{(2)}=1}^{\text{ill. cond. 1}} & \cdots \\
	  \vdots                                 & \vdots                                 & \ddots \\
	\end{bmatrix}
	\begin{bmatrix}
		\chi_{\text{eee},xxx}^{(2)}\\
		\chi_{\text{eee},xxy}^{(2)}\\
		\vdots
	\end{bmatrix}\\
	=
	\begin{bmatrix}
		A_{1,2}\rvert_{\text{exp.}}^{\text{ill. cond. 1}} \\
		A_{1,4}\rvert_{\text{exp.}}^{\text{ill. cond. 1}} \\
		A_{n,1}\rvert_{\text{exp.}}^{\text{ill. cond. 1}} \\
		A_{n,3}\rvert_{\text{exp.}}^{\text{ill. cond. 1}} \\
		\vdots
	\end{bmatrix},
\end{multline}
where the matrix contains the outgoing waves at $\omega_3$ calculated using the known linear properties and supposing that each of the 216 nonlinear susceptibilities is separately equal to $1$. Each set of four rows corresponds to the four outgoing waves for one illumination condition. Each column corresponds to one nonlinear susceptibility. The vector on the left side contains the unknown nonlinear susceptibilities while the vector on the right contains the amplitude of the outgoing modes measured experimentally or simulated for the same 54 illumination conditions.

To verify that a set of illumination conditions are independent, one can calculate the matrix and verify that it is not singular. To ensure stability of the numerical solution of the linear system of equations, the matrix should also have the smallest possible condition number. In our tests, we have found that using random illumination conditions provides a good condition number. However, such illumination conditions are extremely impractical for experiments, and difficult to implement even for simulations. To simplify the setup of the experiment or simulations, it is possible to limit ourselves to a few angles of incidence, and to TE and TM applied waves. Table~\ref{TabIllumination} shows one possible set of illumination conditions that was used for the example below.

\begin{table}
	\begin{tabular}{ccccccccc}
		\hline\hline
		\multicolumn{9}{c}{Mode combinations} \\
		\hline
		\multicolumn{4}{c}{$\omega_1$}                & & \multicolumn{4}{c}{$\omega_2$} \\
		\cline{1-4}                                       \cline{6-9}
		$A_{1,1}$ & $A_{1,3}$ & $A_{n,2}$ & $A_{n,4}$ & & $A_{1,1}$ & $A_{1,3}$ & $A_{n,2}$ & $A_{n,4}$ \\
		    (V/m) &     (V/m) &     (V/m) &     (V/m) & &     (V/m) &     (V/m) &     (V/m) &     (V/m) \\
		\hline
		        1 &         0 &         1 &         0 & &         1 &         0 &         1 &         0 \\
		        1 &         0 &         1 &         0 & &         0 &         1 &         0 &         1 \\
		        0 &         1 &         0 &         1 & &         1 &         0 &         1 &         0 \\
		        0 &         1 &         0 &         1 & &         0 &         1 &         0 &         1 \\
		        1 &         0 &        -1 &         0 & &         1 &         0 &        -1 &         0 \\
		        0 &         1 &         0 &        -1 & &         0 &         1 &         0 &        -1 \\
		\hline\hline
	\end{tabular}\\
	\vspace{0.05in}
 	\begin{tabular}{ccccc}
		\hline\hline
 		\multicolumn{5}{c}{Angle of incidence combinations} \\
		\hline
		\multicolumn{2}{c}{$\omega_1$} & & \multicolumn{2}{c}{$\omega_2$} \\
		\cline{1-2}                        \cline{4-5}
		    $\theta_x$ &    $\theta_y$ & &     $\theta_x$ &    $\theta_y$ \\
		     (degrees) &     (degrees) & &      (degrees) &     (degrees) \\
		\hline
		             0 &             0 & &              0 &             0 \\
		             0 &             0 & &              0 &            30 \\
		             0 &             0 & &             30 &             0 \\
		             0 &            30 & &              0 &             0 \\
		             0 &           -30 & &              0 &           -30 \\
		             0 &            30 & &            -30 &             0 \\
	              30 &             0 & &              0 &             0 \\
	             -30 &             0 & &              0 &            30 \\
	              30 &             0 & &             30 &             0 \\
		\hline\hline
	\end{tabular}
	\caption{A set of 54 independent illumination conditions are obtained by taking all combinations of applied modes (top) and of angles of incidence (bottom).}
	\label{TabIllumination}
\end{table}

In summary, to perform retrieval of the nonlinear susceptibilities of a metamaterial, one should: (1) determine the linear properties of the metamaterial at all frequencies of interest using linear retrieval approaches~\cite{Smith2002Determination, Chen2004Robust, Chen2005Retrieval}; (2) perform a series of independent experiments or simulations (54 in the case of SFG), and measure the nonlinearly generated waves in both modes in media $1$ and $n$; (3) build the matrix by performing a series of calculations using the known linear properties and assuming that each nonlinear susceptibility term is independently equal to 1; and (4) solve the linear system of equations created by the matrix of all forward calculations and the measurements.

\section{Example: varactor-loaded split ring resonator}

\begin{figure}
	\includegraphics{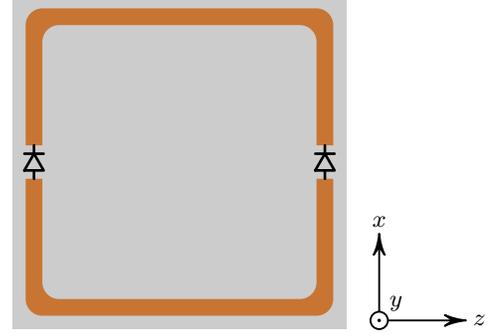}
	\caption{(Color online) The VLSRR considered in the example. The unit cell is cubic with a side of $1\,\text{cm}$. The substrate is FR4 with a thickness of $254\,\mu\text{m}$, covered with $17\,\mu\text{m}$ of copper; the external dimension of the ring is $9.2\,\text{mm}$ and its linewidth is $0.5\,\text{mm}$. Varactor diodes are inserted in the $1\,\text{mm}$ gaps (see text for details). The propagation direction is $z$ while the metamaterial is considered infinite in the $x$ and $y$ directions.}
	\label{FigVLSRR}
\end{figure}

Let us now apply the method we propose to the case of the varactor-loaded split ring resonator (VLSRR). VLSRRs are the canonical example of a nonlinear metamaterial~\cite{Shadrivov2008Nonlinear,Wang2008Nonlinear}. The SRR concentrates electric fields in the small volume of its gaps. If a nonlinear material or component is present in the gap, its nonlinear properties are amplified. We use the VLSRR shown in Fig.~\ref{FigVLSRR}, which has the same dimensions and material properties as those used in Ref.~\onlinecite{Larouche2015Constitutive}. Varactors are included in both gaps and they are oriented in the same direction. When the VLSRR is exposed to electromagnetic waves, the varactors naturally reverse bias. They can therefore be simulated using a simplified model consisting of a $2.35\,\mathrm{pF}$ capacitor in series with a $2.5\,\Omega$ resistor.

\begin{figure}
	\includegraphics{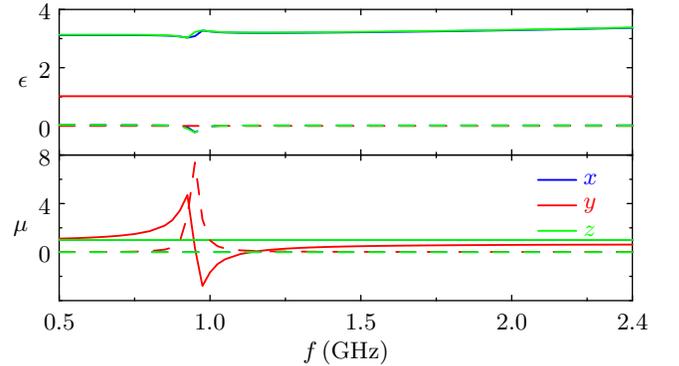}
	\caption{(Color online) Linear properties of the VLSRR of Fig.~\ref{FigVLSRR}. The real part (continuous lines) and imaginary part (dashed lines) of the permittivity (top) and permeability (bottom) are all diagonal.}
	\label{FigSRRLinearProperties}
\end{figure}

The electromagnetic response of the VLSRR was simulated using Comsol~\cite{Comsol52a}. First, a series of time-harmonic linear simulations was performed illuminating the SRR from various directions and with various polarizations. The standard linear retrieval approach was used to determine the linear properties, shown in Fig.~\ref{FigSRRLinearProperties}. As expected, the dominant response of the SRR is a magnetic resonance in the $y$ axis, which occurs around $0.9\,\text{GHz}$. There is no magnetic response in the two other axes. The SRR also has a non-resonant electric response for fields polarized in the plane of the SRR.

Next, the 54 nonlinear simulations described in Tab.~\ref{TabIllumination} were performed to determine the outgoing fields at frequency $f_3 = \omega_3/2\pi$. In those simulations, we varied $f_1 = \omega_1/2\pi$ between $0.5\,\text{GHz}$ and $1.5\,\text{GHz}$, while $f_2 = \omega_2/2\pi$ was kept constant at $0.9\,\text{GHz}$. The simulations involved three time-harmonic simulations at the three frequencies involved. At each frequency, two circuit models, one for each varactor, were connected inside of the gaps. The nonlinear coupling between the three frequencies occurs in the varactor, whose capacitance is nonlinear. To account for this, we assume that the potentials on the capacitor at $f_1$ and $f_2$, $V_1$ and $V_2$, generate a potential at $f_3$ given by $V_3 = a_2V_1V_2$ where $a_2 = 0.2667\,\mathrm{V}^{-1}$ comes from a power series expansion of the varactor capacitance~\cite{Poutrina2010Analysis}.

\begin{figure*}
	\includegraphics{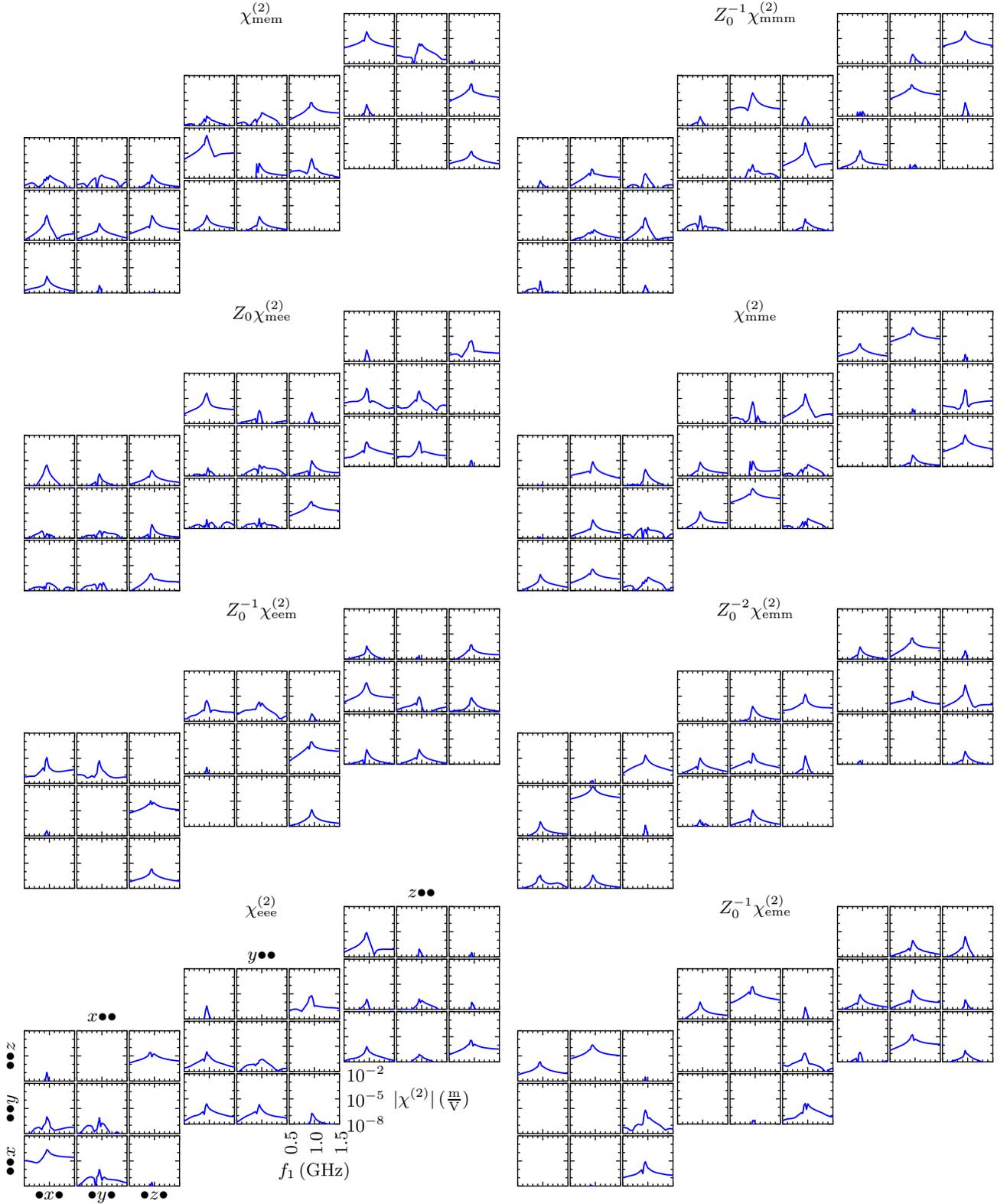}
	\caption{(Color online) Norm of the second order nonlinear susceptibilities of the VLSRR. To put all values on the same scale, nonlinear susceptibilities involving magnetic terms are normalized using the impedance of vacuum, $Z_0$. For each type of nonlinearity, the results are presented in three slices, corresponding to the orientation of the field at $f_3$. In each slice, columns and rows correspond to the orientation of the fields at $f_1$ and $f_2$, respectively. All subplots share the same axes.}
	\label{FigSRRNonlinearProperties}
\end{figure*}

We then used Matlab~\cite{Matlab2016b} to generate the retrieval matrix, and solve the linear system of equations to retrieve all 216 nonlinear susceptibility terms shown in Fig.~\ref{FigSRRNonlinearProperties}. We determined the complex values for the nonlinear susceptibilities, but for simplicity only show their norm. The range of simulations we performed includes second harmonic generation ($f_1 = f_2 = 0.9\,\mathrm{GHz}$). In that case, the fields at $f_1$ and $f_2$ and indistinguishable; to avoid an underdetermined system of equations, we assumed that $\chi_{abc,\alpha\beta\gamma}^{(2)} = \chi_{acb,\alpha\gamma\beta}^{(2)}$.

The main result that can be observed in Fig.~\ref{FigSRRNonlinearProperties} is that $\chi^{(2)}_{emm,xyy}$ is about one order of magnitude larger than any other term. This is not surprising; the applied waves at $f_1$ and $f_2$ are both close to the magnetic resonance frequency of the element, such that a magnetic field in the $y$ axis couples well in the VLSRR. Since the two diodes are oriented in the same direction, they generate potentials at $f_3$ that are also in the same direction. This symmetry corresponds to that of an electric field in the $x$ axis.

The term $\chi^{(2)}_{eee,xxx}$ is also supported by the geometry. Electric fields oriented along the $x$ axis at $f_1$ and $f_2$ polarize the two varactors, which generate an electric field in the $x$ direction at $f_3$. However, when the VLSRR is excited by an electric field, most of field concentrates between adjacent rings, rather than in the gaps, such that this nonlinearity is significantly smaller than $\chi^{(2)}_{emm,xyy}$. Similar arguments can be made for the terms $\chi^{(2)}_{eem,xxy}$ and $\chi^{(2)}_{eme,xyx}$.

The retrieval indicates that many other nonlinearities are present which do not seem to be supported by the geometry. They can all be explained by spatial dispersion~\cite{Rose2012Nonlinear,Gorlach2016Nonlocal}. Spatial dispersion occurs because the metamaterial elements, which are discrete, are replaced by a homogeneous layer of finite thickness where the nonlinearity is distributed. The magnitude of this effect is related to the ratio between the size of the unit cell and all the wavelengths involved. At the resonance frequency, the refractive index of the VLSRR is maximum with a value of about four. The wavelength inside of the metamaterial is about $8\,\text{cm}$, less than an order of magnitude larger than the unit cell. Therefore, spatial dispersion shows mainly at that frequency.

\section{Conclusion}

We have proposed a nonlinear transfer matrix appropriate for bi-anisotropic materials with any combination of nonlinearities and showed how this approach can be used to retrieve the effective nonlinear susceptibilities of a metamaterial.

For simplicity, we have demonstrated the approach using SFG, a second order process. However, our approach generalizes easily to a process of any order, with a rapidly increasing number of terms. In general, for a $n$th order process, there are $2^{n+1}$ nonlinear susceptibility tensors, each containing $3^{n+1}$ terms, for a total of $6^{n+1}$ independent terms. For a third order process, for example, there are 1296 elements.

We hope that the application of our method will encourage the development of complex nonlinear metamaterials with many nonlinear susceptibilities enabling new exciting applications.

\begin{acknowledgements}

Approved for Public Release: NG17-2096, 10/23/17

\end{acknowledgements}

%

\end{document}